# Robust Multi-echo GRE Phase processing using a unity rank enforced complex exponential model

Joseph Suresh Paul, Sreekanth Madhusoodhanan
Medical Image Computing and Signal Processing Laboratory, Indian Institute of Information Technology and Management- Kerala, Trivandrum – 695581, India
j.paul@iiitmk.ac.in

**Abstract**

**Purpose:** Develop a processing scheme for Gradient Echo (GRE) phase to enable restoration of susceptibility-related (SuR) features in regions affected by imperfect phase unwrapping, background suppression and low signal-to-noise ratio (SNR) due to phase dispersion.

**Theory and Methods:** The predictable components sampled across the echo dimension in a multi-echo GRE sequence are recovered by rank minimizing a Hankel matrix formed using the complex exponential of the background suppressed phase. To estimate the single frequency component that relates to the susceptibility induced field, it is required to maintain consistency with the measured phase after background suppression, penalized by a unity rank approximation (URA) prior. This is formulated as an optimization problem, implemented using the alternating direction method of multiplier (ADMM).

**Results:** With in vivo multi-echo GRE data, the magnitude susceptibility weighted image (SWI) reconstructed using URA prior shows additional venous structures that are obscured due to phase dispersion and noise in regions subject to remnant non-local field variations. The performance is compared with the susceptibility map weighted imaging (SMWI) and the standard SWI. It is also shown using numerical simulation that quantitative susceptibility map (QSM) computed from the reconstructed phase exhibits reduced artifacts and quantification error. In vivo experiments reveal iron depositions in insular, motor cortex and superior frontal gyrus that are not identified in standard QSM.

**Conclusion:** URA processed GRE phase is less sensitive to imperfections in the phase pre-processing techniques, and thereby enable robust estimation of SWI and QSM.

# 1. INTRODUCTION

Of late, susceptibility weighted imaging (SWI)(1-4) and quantitative susceptibility mapping (QSM)(5-10) has increasingly gained importance as a clinical tool for visualization of fine venous structures, bleeds and mineral depositions. Since the phase in gradient echo (GRE) acquisition is directly related to the local field variations caused by the susceptibility sources (11, 12), it is used to enhance the susceptibility-related (SuR) contrast of magnitude SWI, and estimate susceptibility maps. Contrast enhancement has been lately achieved using both non-linear phase filters(13), and weighting using the susceptibility map(14). Moreover, as the phase wraps and signal-to-noise ratio (SNR) vary widely with echo time, a multi-echo acquisition has always been beneficial for GRE to reduce the spatial distortion and increase the SNR (15, 16). More significantly, this also allows estimation of a signal model at each voxel after the pre-processing steps are applied to unwrap the phase and remove the macroscopic background field (17-19).Imperfections in the standard phase pre-processing are known to result in loss of spatial resolution and SuR information(20, 21). Therefore, a major challenge is to recover the lost information in regions with high phase wraps and macroscopic field inhomogeneities. Of particular relevance is the apparent non-linear temporal phase evolution and TE-dependence of frequency and QSM measurements in and around focal susceptibility sources(22). Imperfections of phase pre-processing in the presence of noise can also introduce additional errors in the estimation of phase contributed by the local field inhomogeneities.

In the case of SWI and QSM, one needs to correct the local phase after suppressing the effect of macroscopic field variations. Ideally, the complex exponential of the resultant background suppressed phase should represent the frequency-shift corresponding to the susceptibility induced field. This can be estimated by fitting the complex exponential signals along the temporal dimension with a single frequency at each voxel. The fitting is obtained using Kronecker's theorem according to which, a Hankel matrix generated from a uniformly sampled temporal signal has rank *P* if and only if they can be fitted with a complex exponential having *P* frequency components(23, 24). Since this method is based on uniformly sampled complex exponential signal, it fits very well with the multi echo SWI and QSM acquisition conventionally used in the clinical applications. Other than the dual echo, most multi-echo acquisitions are uniformly sampled in the practical condition. Since it is required to fit the complex exponential signals with a single frequency component in the current context, this amounts to applying a unity rank approximation (URA) of the Hankel matrix. Thus the problem of estimating the phase without making it deviate too much from the measured phase after background suppression along with the URA prior, presents this as an optimization problem, easily implementable using the alternating direction method of multiplier (ADMM) with variable splitting (25, 26).

## 2. THEORY

### 2.1 Phase pre-processing for SWI and QSM

The phase of a multi-echo GRE sequence is given by

$$\psi_r^l = 2\pi\gamma \cdot \Delta B_r \cdot (\text{TE}^0 + l\Delta\text{TE}) + \Omega_r^l + 2\pi\epsilon_r^l, \qquad [1]$$

where $\gamma$ is the gyromagnetic ratio (MHz/T), $\Delta B_r$ is the field perturbation (T) at location $r$, $\text{TE}^0$ is the initial echo time, $\Delta\text{TE}$ is the echo spacing, $l$ is the echo index, $l = 0,1,2 \ldots, L-1$, $L$ is the number of echoes, $\Omega^l$ is the phase contribution from additive noise term, and integer $\epsilon_r^l$ is the phase wrapping term whose value will be non-zero when the sum of first two terms falls outside the $[-\pi, \pi)$. Since the phase information is directly proportional to the susceptibility variations and echo time, phase wraps also increase with the echo time. Therefore, to extract the local SuR field information, phase unwrapping (27-29) followed by background suppression is usually performed (13, 30-32).

In 3D GRE with sufficiently long TE, intra voxel dephasing can introduce signal losses particularly at low imaging resolution (slice thickness is much greater than the in-plane resolution) (33), which is typically the case in SWI clinical imaging and QSM. The resultant signal model for the magnitude intensities along temporal dimension loses the linear predictability due to the sinc function modulation of the magnitudes (33, 34). Thus in situations where the echoes are few in number, the required number of samples to deduce the signal model for the magnitude component (such as the sinc function) may not be adequate. This makes it difficult to apply any procedure for correction of magnitude intensity by estimating the extent of intra voxel dephasing. Therefore, introduction of the magnitude intensity into the model fitting can significantly influence the phase estimate. This effect is more pronounced particularly in regions with signal dropout and also those having low intensities that sometimes overlap with areas containing the venous structures to be restored.

**2.2 Low Rankness in Multi-echo data**

The measured signal from a GRE sequence at echo time TE is given by (35, 36)

$$g_r^l = \rho_r^l \cdot \exp\{i\psi_r^l\}, \qquad [2]$$

where $\rho_r^l$ is the signal intensity measured at location $r$ for echo index $l$. In acquisition with more than one echo, the reconstructed images are denoted as $g^0, g^1, \cdots, g^{L-1}$. As the phase wraps become more severe at higher echo times, one needs to apply phase unwrapping and background suppression to the phase component of $g$, prior to reconstruction with URA prior. Let $\varphi^l$ be the phase image obtained after unwrapping and background suppression of $\psi^l$. Since Kronecker's theorem is valid only for predictable signal components and the magnitude intensity evolution along the temporal dimension does not constitute a predictable signal due to the effect of intra voxel dephasing, we only consider the complex exponential of the background suppressed phase (with unit magnitude) for generating the initial Hankel matrix. For a given location $r$, the collection of unit magnitude complex exponential signals at each TE is modelled as a uniformly sampled signal with a single frequency component. This is given by

$$x_r = \begin{bmatrix} e^{i\varphi_r^0} \\ e^{i\varphi_r^1} \\ \vdots \\ e^{i\varphi_r^{L-1}} \end{bmatrix} = \begin{bmatrix} e^{i2\pi\gamma\cdot\Delta B_r\cdot(\text{TE}^0)+\hat{\Omega}_r^1} \\ e^{i2\pi\gamma\cdot\Delta B_r\cdot(\text{TE}^0+\Delta\text{TE})+\hat{\Omega}_r^2} \\ \vdots \\ e^{i2\pi\gamma\cdot\Delta B_r\cdot(\text{TE}^0+(L-1)\Delta\text{TE})+\hat{\Omega}_r^L} \end{bmatrix}, \quad [3]$$

where $x_r$ represents a vector with elements consisting of the exponentiated background suppressed phase at each location. To permit further analysis on $x_r$, a common approach is to map the one dimensional signal $x_r$ to a multi-dimensional structured Hankel matrix $H$. Using the window length ($w$) as the parameter for mapping such that $2 \leq w \leq L(37)$, the structured Hankel matrix $H_r$ is constructed as

$$H_r = \begin{bmatrix} e^{i\varphi_r^0} & e^{i\varphi_r^1} & \cdots & e^{i\varphi_r^{w-1}} \\ e^{i\varphi_r^1} & e^{i\varphi_r^2} & \cdots & e^{i\varphi_r^w} \\ \vdots & \vdots & \ddots & \vdots \\ e^{i\varphi_r^{L-w}} & e^{i\varphi_r^{L-w+1}} & \cdots & e^{i\varphi_r^{L-1}} \end{bmatrix}. \quad [4]$$

If the window length is chosen to be $w = (L+1)/2$, the Hankel matrix would be square ($w \times w$), so that singular value decomposition (SVD) for rank minimization can be performed in minimum number of flops (38, 39). Since the elements of the Hankel matrix H are uniformly sampled points with single frequency component, it can be considered as a matrix with rank $R=1$.

In practical conditions, due to the presence of noise in the measured signal, the small singular components of the Hankel matrix H may not be zero. Therefore singular values of H can be truncated to obtain a minimum variance estimate. Since the matrix obtained after singular value thresholding loses the Hankel structure, the reconstructed signal can be recovered from the rank reduced H by averaging across its anti-diagonal elements. This step helps in iterative filtering of the signal model (37). The processed multi-echo phase image $\varphi$ is recovered as the imaginary part of the complex logarithm applied to the elements of H. Alternatively, $\varphi$ can also be recovered using the principle angle of the elements of H, with the idea that the phase values are limited in the range of $[-\pi, \pi)$. To maintain consistency with the measured phase after background suppression, a fidelity term is added along with the rank prior. The estimate of the underlying true phase representative of the SuR information can hence be obtained as the solution to the optimization problem:

$$\underset{y_r}{\operatorname{argmin}} \|x_r - y_r\|_2^2, \quad subject\ to\ rank\ (\mathrm{H}_r) = 1, \qquad [5]$$

where $x_r$ denotes the observed complex exponential signal at location $r$, with unit magnitude. Although $x_r$ is of unit magnitude, the recovered signal $y_r$ may contain a magnitude component $< 1$, that is not relevant to the desired phase information. Our numerical observations reveal that the magnitudes after cross-diagonal averaging in successive iterations are very close to unity and do not affect the reconstructed phase image.

Hereafter, we drop the pixel subscript $r$ with the understanding that the remaining analysis applies separately to each voxel in the image. To enable solution using ADMM, the above rank minimization problem can be restated as

$$\underset{H,u}{\text{argmin}}\, \frac{1}{2}\|u\|_2^2 + \lambda \mathbb{R}(H), \; subject\; to$$
$$H(j,k) + u(j+k) = x(j+k); \; 0 \leq j,k \leq M, \quad [6]$$

where $u = x - y$, $M = \frac{L-1}{2}$, $\lambda$ is a regularization parameter that controls the balance between the fidelity and prior terms. In Eq. [6], the prior term $\mathbb{R}(\cdot)$ is the indicator function defined as

$$\mathbb{R}(H) = \begin{cases} 0 & if\; rank\,(H) = 1 \\ \infty & otherwise \end{cases}. \quad [7]$$

Since the cost function consist of two terms, each depending only on one of the variables H and $u$, the problem formulated in Eq.[6] is well suited to be addressed using ADMM. For the rank minimization applied to a Hankel matrix $H\epsilon\mathbb{C}^{M\times M}$ with complex exponentials of magnitudes $\leq 1$ as its elements, the singular values are always bounded, ie. $\|H\| \leq M$, where $\|\cdot\|$ denotes the spectral norm. As the rank minimization is known to be computationally intractable (NP-hard) (40), the rank reduction can be made effective by choosing the convex hull of the function $\phi(H) = rank\,(H)$ over the set $\mathcal{C} = \{H\epsilon\mathbb{C}^{M\times M}; \|H\| \leq 1\}$ as $\hat{\phi}(H) = \|H\|_* = \sum_{i=1}^{M} \sigma_i$ (c.f theorem 6.5.2 in (41)). Consequently, this requires an initial scaled version $H \triangleq \frac{1}{M}H$ to ensure $\|H\| \leq 1$. Although the phase expressed as complex exponential in the fidelity term facilitates the ADMM implementation, it makes the cost function non-convex due to the periodicity of the complex exponential(42). Based on previously reported fact that the background suppressed phase will have an intensity $|\varphi(r)| \leq \pi$(43),the cost function can be treated to be convex in the current context. This additionally favours the convergence to a global minimum. For obtaining the ADMM solution, it is required to consider the augmented Lagrangian function of Eq.[6]:

$$\mathcal{L}(H, u, z) = \lambda \mathbb{R}(H) + \frac{1}{2}\|u\|_2^2 + \sum_{j,k=0}^{M} z(j,k) \cdot [H(j,k) + u(j+k) - x(j+k)]$$

$$+ \sum_{j,k=0}^{M} \frac{\delta}{2} |H(j,k) + u(j+k) - x(j+k)|^2, \quad [8]$$

where z is the Lagrange multiplier of the linear constraint, and $\delta$ is the penalty parameter. The ADMM iterative steps from iteration $q$ to $q+1$, involve solving the following subproblems:

$$H^{q+1} = \underset{H}{\mathrm{argmin}}\, \lambda \mathbb{R}(H) + \sum_{j,k=0}^{M} \frac{\delta}{2} \left| H(j,k) + u^q(j+k) - x(j+k) + \frac{1}{\delta} z^q(j,k) \right|^2 \quad [9]$$

$$u^{q+1} = \underset{u}{\mathrm{argmin}}\, \frac{1}{2}\|u\|_2^2$$
$$+ \sum_{j,k=0}^{M} \frac{\delta}{2} \left| H^{q+1}(j,k) + u(j+k) - x(j+k) + \frac{z^q(j,k)}{\delta} \right|^2 \quad [10]$$

$$z^{q+1}(j,k) = z^q(j,k) + \delta[H^{q+1}(j,k) + u^{q+1}(j+k) - x(j+k)]$$
$$for\ 0 \leq j,k \leq M \quad [11]$$

The ADMM iterative steps for solving the above problem are sequentially implemented using the step that involves the proximal approximation involving the unity rank condition. The primal variable update is obtained using the proximal operator for nuclear norm functional with fixed rank approximation, applied to the least squares solution of the subproblem (44)

$$\widehat{H}^{q+1}(j,k) = x(j+k) - u^q(j+k) - \frac{1}{\delta}(z^q(j,k)) \quad [12]$$

The updated $H^{q+1}$ is then obtained using the proximal approximation followed by projecting the resultant matrix onto the space of Hankel matrices by application of the cross-diagonal averaging operator. This is implemented using

$$H^{q+1} = \mathcal{D}_{\sigma_{1min} > \frac{\lambda}{\delta} \geq \sigma_{2max}}(\widehat{H}^{q+1}) \quad [13]$$

where $\mathcal{D}$ represents the singular value thresholding (SVT) followed by cross-diagonal averaging, with the threshold range as determined by the unity rank. $\sigma_{1min}$ and $\sigma_{2max}$ denote the minimum primary singular value and maximal secondary singular value estimated from the collection of Hankel matrices corresponding to each voxel within the ROI.

Based on the signal model that we have considered, the susceptibility induced phase after background suppression, essentially consist of a single frequency component at each voxel. Hence as per the kronecker's theorem, the Hankel structure should have an effective rank one. Nevertheless, for cases where the singular values repeat, it evidently points out that there are two or more frequency components having the same amplitude. Thus unity rank enforcement for such cases can introduce errors. Further details of the ADMM implementation are provided in the supplementary material.

To generate a symmetric square Hankel matrix for the case where $L$ is even, the vector $x_r$ can be split into two vectors $^1x_r$ and $^2x_r$ using first and last $(L-1)$ elements of $x_r$. The two Hankel matrices $^1H$ and $^2H$, each of size $\frac{L}{2} \times \frac{L}{2}$, can be constructed by choosing the window length $w = \frac{L}{2}$(45). To compute $^1u$ and $^2u$, the minimization procedure can then be performed separately on $^1x_r$ and $^2x_r$. The first and last element of the solution $u$ can be directly obtained from the first element of $^1u$, and the last element of $^2u$. The rest of the elements of $u$ can be estimated by averaging the last $(L-2)$ elements in $^1u$ and the first $(L-2)$ elements in $^2u$.

## 3. METHODS

### 3.1 In vivo Experiments

SWI and QSM datasets were acquired using a 3D multi-echo GRE sequence on a 3.0 T MRI scanner (Discovery MR750W, GE, Milwaukee, WI) with a 12 channel head array coil. Healthy volunteers who did not have a medical history of neurological disease were selected for this study. This study was approved by the institutional ethics committee and written informed consent was obtained from all participants. A total of 12 volunteers with an average age of 40 were scanned for SWI and QSM.

SWI data acquisition and processing

For in vivo experiments, two sets of fully sampled multi-echo flow compensated 3D GRE data were acquired. The first set, consisting of three echoes, was acquired with TE ranging from 20 to 29.36 ms, with an interval of 4.68 ms, repetition time (TR) = 34 ms, flip angle (FA) = 20°, slice thickness 2.4 mm, acquisition matrix 384 × 288× 54, bandwidth (BW) = 325.52 Hz/pixel and field of view (FOV) = 260 × 260×130 mm³. The second data set was a five echo data acquired at an initial TE of 5.7ms and echo spacing of 5.9 ms, TR = 61.9 ms, FA = 20°, slice thickness 2.0 mm, acquisition matrix 384 × 202× 64, BW = 300 Hz/pixel and FOV = 260 × 260×130 mm$^3$.The fully sampled multi-channel data was first channel combined into a single complex data using Roemer's method(46). This was followed by Laplacian based fast unwrapping (28), background suppression using projection onto dipole fields (PDF)(31) and reconstruction using URA prior, preceding phase mask generation. As the GE scanner follows the right handed convention, the negative values of the processed phase were used to generate the phase mask. Finally, the contrast enhanced magnitude images were generated by multiplying the original magnitude image with the phase mask. In order to maximize the contrast-to-noise ratio (CNR), the number of phase mask multiplications was limited to 4(43, 47). For better visualization of venous structures, minimum intensity projection (mIP) was obtained using 4 slices of the magnitude SWI.

QSM data acquisition and processing

The second set of 3D GRE sequence was used for the QSM experiments. QSM computation was performed using morphology enabled dipole inversion (MEDI) algorithm (48, 49),following Laplacian based fast unwrapping and background suppressionusing PDF.Decreasing the value of MEDI regularization parameter causes an increase in the number of streaking artifacts. Hence in all our experiments, the MEDI regularization parameter was fixed to 1000(50, 51). Quantitative assessment was performed by measuring the susceptibility values of several brain anatomical structures including Globus-Pallidus (GP), Putamen (PT), Caudate Nucleus (CN), Thalamus (TH), White Matter (WM), Substantia Nigra (SN), Red Nucleus (RN) and Dentate Nucleus (DN). ROIs of these structures were manually selected from slices of susceptibility maps. The mean susceptibility within each ROI was recorded for further analysis.

**3.2 Processing pipeline**

In standard SWI processing, a homodyne filter is typically used for background suppression without the need for performing a prior phase unwrapping step. With homodyne filter, residual phase wraps will occur when the filter size is too small, while the image contrast degrades when the filter size is too large. Therefore, maintaining image contrast with simultaneous removal of wraps require phase unwrapping (13). However, this is accomplished at the cost of additional computation. Therefore, for the present study, background suppression was performed after spatial phase unwrapping using techniques such as PDF, SHARP or LBV (30-32).

For uniformity, we have used PDF for all in vivo experiments. For reconstruction with URA prior at each location, the unit magnitude complex exponential signals along the temporal direction were used to construct a square Hankel matrix. Singular value thresholding was then performed on the Hankel matrix by retaining the highest singular value and principal singular vectors. To restore the Hankel structure, each matrix element was replaced by the mean of its cross-diagonal elements. By iteratively continuing the rank minimization together with imposition of data fidelity constraint, empirical convergence was observed in terms of the relative $l_2$-norm error between phase images of successive iterations. The phase image generated after attaining convergence was used for either SWI or QSM processing. The proposed processing pipeline is shown in Fig.1.

## 3.3 Choice of free parameters

One of the main parameter to control the reconstruction performance is the regularization parameter ($\lambda$) used to update the complex exponential. Both the $\lambda$ and ADMM penalty parameter $\delta$ influences the reconstruction performance in similar lines to that reported by Ghadimi(52). Our numerical experiments using simulated data reveals dependence of mean-square-error (MSE) on the choice of $\lambda$. The MSE is computed as the mean of the squared absolute difference between the reconstructed and groundtruth phase images at each voxel. The figure below shows MSE versus $\lambda$ plots at two noise levels. This influence of regularization mainly arises from the dependence of the threshold used for the SVT operation. Since the singular value soft thresholding parameter lies in the range $\sigma_{1min} > \tau \geq \sigma_{2max}$ for unity rank approximation, the value of optimal $\lambda$ lies in the range $\delta\sigma_{1min} > \lambda^* \geq \delta\sigma_{2max}$. Here $\sigma_{2max}$ denotes the maximum secondary singular value of the Hankel matrices constructed at each voxel.

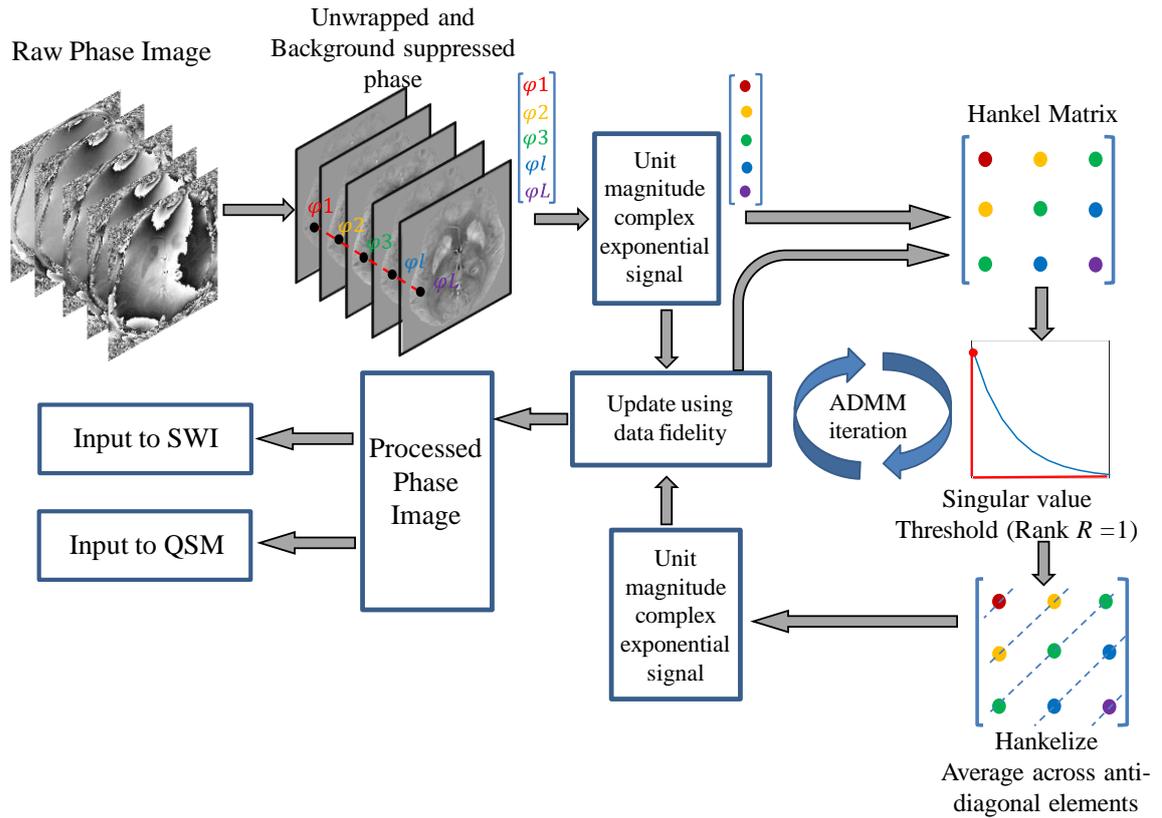

**Figure 1**: Schematic block diagram of GRE phase processing pipeline.

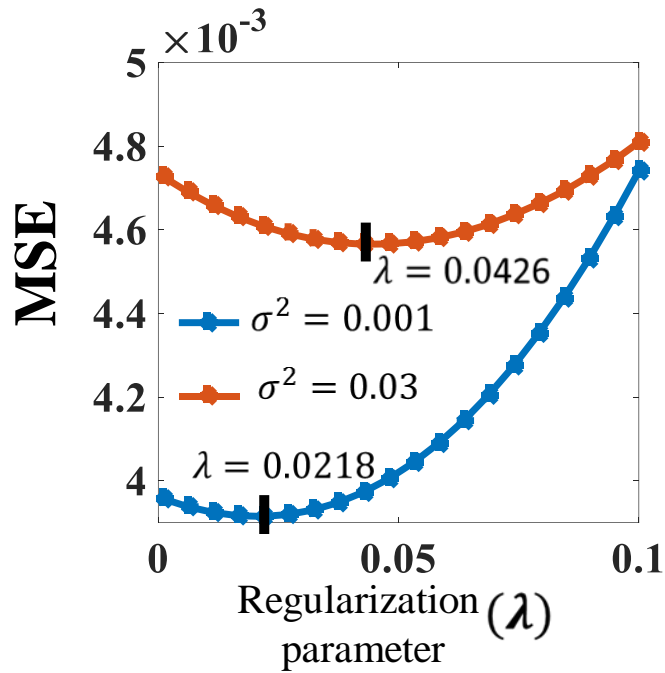

Figure 2: Dependence of MSE on λ. Blue and red curves correspond to noise levels $\sigma^2 = 0.001$ and $0.03$.

The step-size or ADMM penalty parameter $\delta$ mainly determines the convergence rate (25). In the ADMM implementation, this is adaptively chosen to balance between the two terms in the distance from convergence measure defined as the sum of squared $l_2$-norms of the primal and dual residuals(53).

## 4. RESULTS

### 4.1 Simulation experiments

Using the QSM from the combined echoes of one of the volunteer data as the true susceptibility distribution (groundtruth) $\chi(r)$, the field perturbation along the z-direction of the dipolar magnetic field is simulated using (54)

$$B_{dz}(k) = -\left(\frac{\chi(k)B_0}{3}\right)\left(3\frac{k_z^2}{k_x^2 + k_y^2 + k_z^2} - 1\right), \qquad [14]$$

where $\chi(k)$ is the Fourier transform of $\chi(r)$, $k_x, k_y$ and $k_z$ represent the k-space coordinates at position k. These field maps are then used to generate idealized phase data ($\phi$) that accumulates linearly over time. Simulated phase data are generated for echo times $TE1/\Delta TE/TE5 = 5.6/5.9/29.8$ms. Complex data are then synthesized using unit magnitude and the simulated phase. Each magnitude and phase pair is then converted into real and imaginary images, to each of which a normally distributed zero mean noise with variance $\sigma^2$ is added to ensure comparable SNR in the simulated data. The noisy wrapped phase data are then generated using

$$\psi = \arctan\left(\frac{Im(exp(i\phi)) + \mathfrak{N}(0,\sigma^2)}{Re(exp(i\phi)) + \mathfrak{N}(0,\sigma^2)}\right) \qquad [15]$$

QSM data is generated after phase unwrapping and background suppression. The top row in Fig. 3 below shows the groundtruth susceptibility maps ($\chi$) for different slices that contain specific regions of interest cited in our in vivo experiments. The wrapped phase for each echo time was computed as the principal angle obtained after noise addition as in Eq. [15]. The noise variance was chosen to be 3% of the maximum amplitude (unity in the present case). QSM maps were then generated following the steps outlined in the processing pipeline.

Second and third rows in Fig. 3 show the susceptibility maps generated without and with URA prior. Red circles highlight regions with artifacts in the susceptibility map. It is observed that application of URA prior eliminates the above artifacts, as indicated by the yellow circles. The bottom two rows show the difference images obtained with respect to the groundtruth, in which the white circles indicate regions with artifacts.

Arrow heads are used to show those regions with significant quantification errors that are reduced in the susceptibility maps obtained with URA prior. An example is a brain slice with basal ganglia structure (first column) for which the difference image obtained using susceptibility maps without URA prior show high intensity values in the caudate nucleus. Susceptibility maps in the second column show the influence of URA prior in the frontal and motor cortex region. It is seen that both the artifact in the motor cortex region and quantification error in the frontal region are simultaneously reduced. In slices with interpeduncular cistern and dentate nucleus as shown in the third and fourth columns, reonstruction with URA prior results in a 15% reduction in the quantification error.

The tolerance is used only as a stopping criterion using the relative error between successive iterates. This is because in simulation experiments, we find that once the relative error falls within the specified tolerance, the MSE measured with reference to the groundtruth do not vary significantly.

**4.2 Application of phase processing to in vivo SWI**

Fig.4 shows sample phase images of a multi-echo SWAN sequence, before and after reconstructing the phase with URA prior (first and second row) using $\lambda = 0.004$. Column-wise panels show the phase images at $TE = 20$ ms, 24.6 ms, 29.3ms, and the echo-combined phase. In the combined phase images, yellow arrows indicate the newly observed transverse veins that are not seen in the unprocessed combined phase image. Furthermore, the improved gray-white matter contrast in the processed phase image enables better visualization of the white matter tracks. The blue arrow in the combined phase points to the white matter track between lentiform nucleus and insular region that is hardly seen in the unprocessed phase image. The empirical convergence of the phase reconstruction scheme is shown using the relative $l_2$-norm error between the phase images estimated in successive iterations in Fig.4.

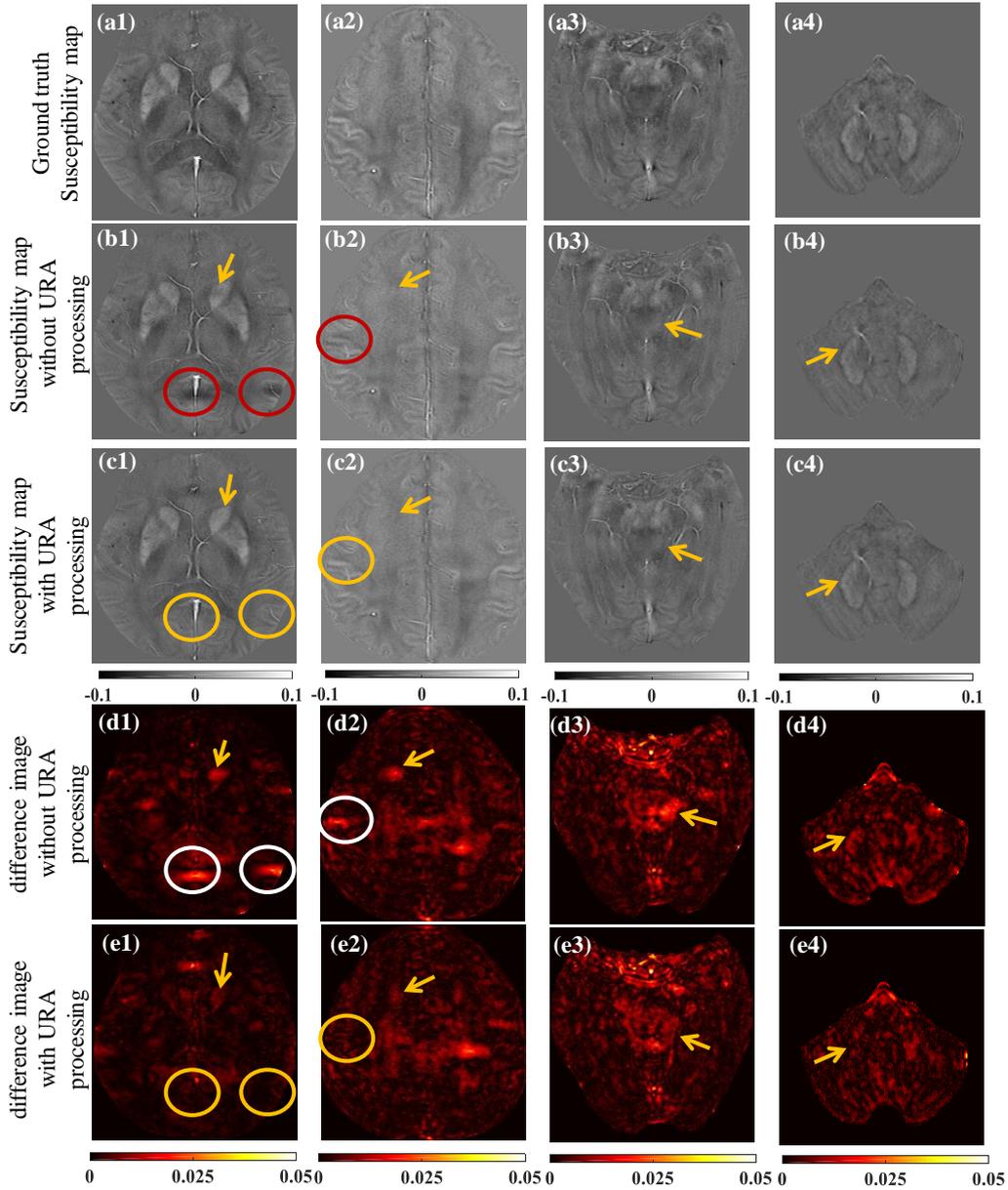

**Figure 3:** The top row shows the groundtruth susceptibility maps for four different slices. Second and third rows show the susceptibility maps generated from phase image processed without and with URA prior. Red circles highlight regions with artifacts in the susceptibility map. Application of URA prior eliminates the above artifacts, as indicated by the yellow circles. The bottom two rows show the difference images obtained with respect to the groundtruth. The in which the white circles indicate regions with artifacts.

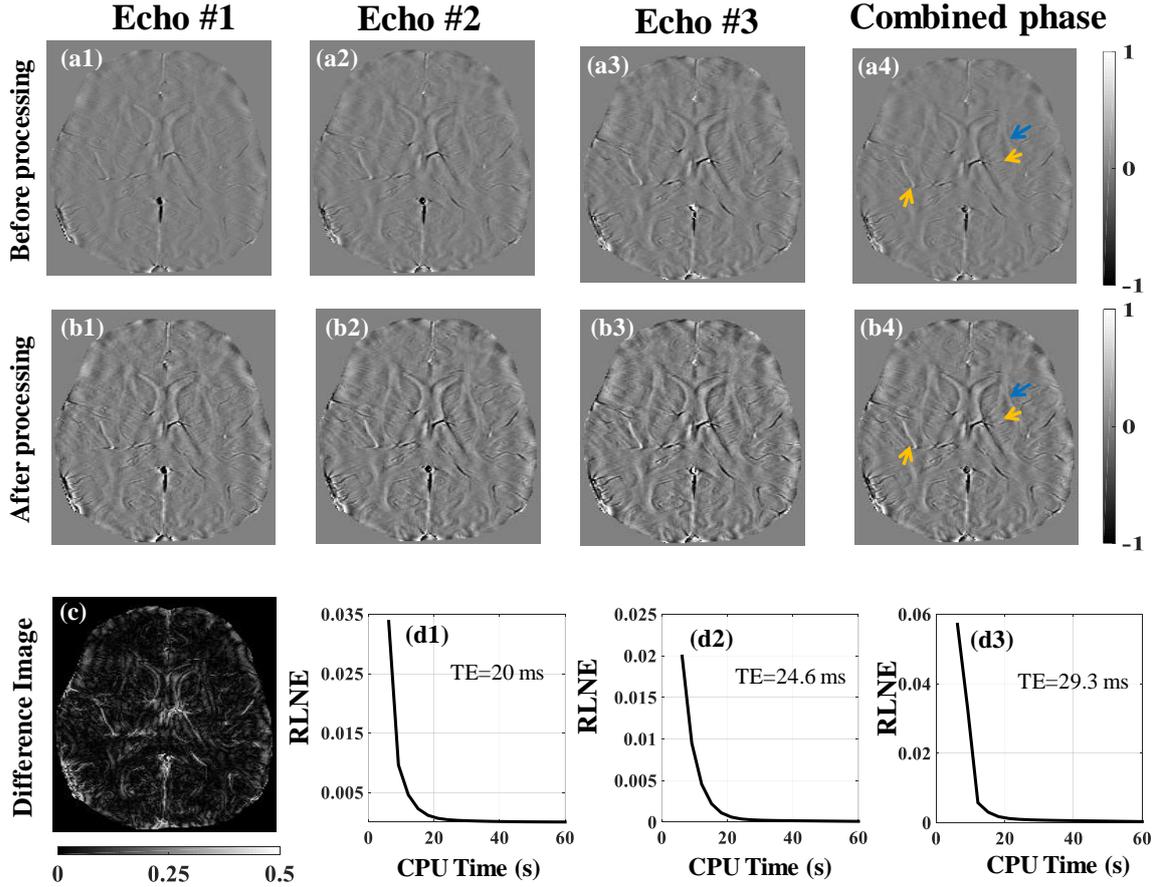

**Figure 4:** Top two panels show the unprocessed phase image and phase image reconstructed with URA prior. Column-wise panels show phase images at $TE = 20$ ms, 24.6 ms, 29.3 ms and the echo-combined phase. Yellow and blue arrows are used to indicate the newly observed transverse veins and white matter tracks in the echo-combined reconstructed phase image. Panel (c) in the bottom row shows the absolute difference map of the phase image before and after processing.(d1)-(d3) show the empirical convergence plots of the reconstructed phase for each echo.

Panel-(a) in Fig.5 shows the ROIs (R1-R4) used for calculating the gray-white matter contrast. Red and yellow squares indicate the gray and white matter regions at each ROI. Bar plots depict the gray-white matter contrast computed from the phase image without and with URA prior. It is observed that the contrast obtained with URA prior is higher in all the ROIs and the mean contrast improvement is found to be two fold.

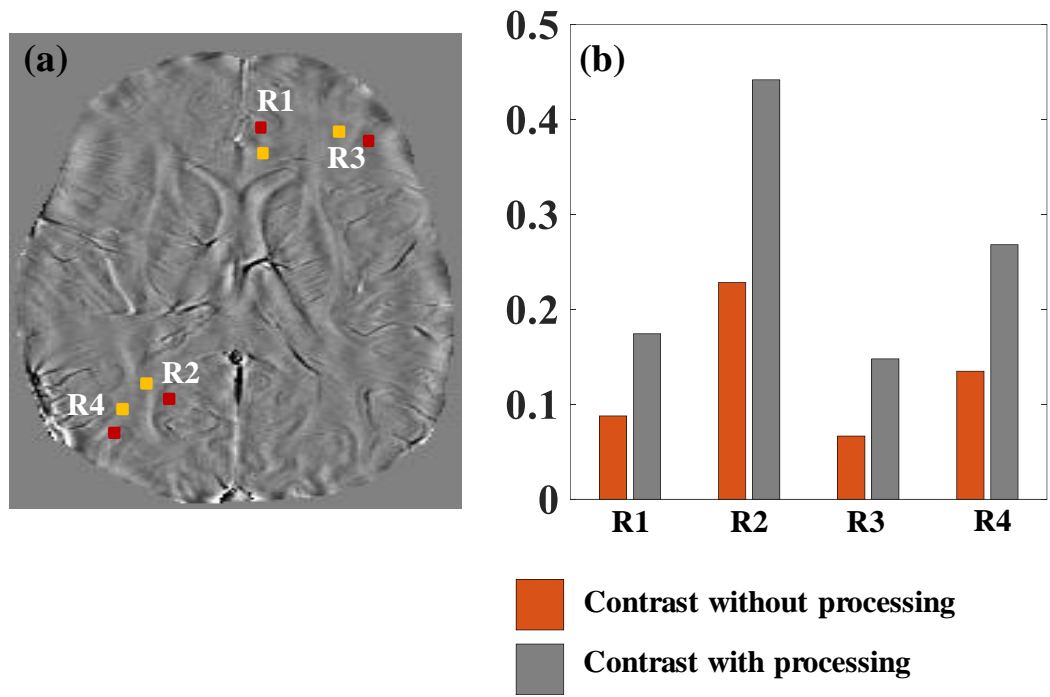

**Figure 5:** Gray-white matter contrast computed from the phase image without and with URA prior. Red and yellow squares indicate the gray and white matter regions at each ROI.

Fig. 6 illustrates the application to SWI processing. The combined phase, phase mask, and magnitude SWI obtained with and without URA prior are shown in the left and right panels, respectively. Red arrows indicate the faint transverse and medullary venous structures that are highlighted only in the magnitude SWI generated using the reconstructed phase images.

A major advantage of phase reconstruction with URA prior in SWI processing is the restoration of obscured information in regions with residual components of macroscopic field inhomogeneity and phase wraps. It is known that the non-local influence from distant magnetization will affect the magnetic field at a nearby voxel. The surrounding field inhomogeneity significantly alters the field and BOLD signal at the region of interest(55).

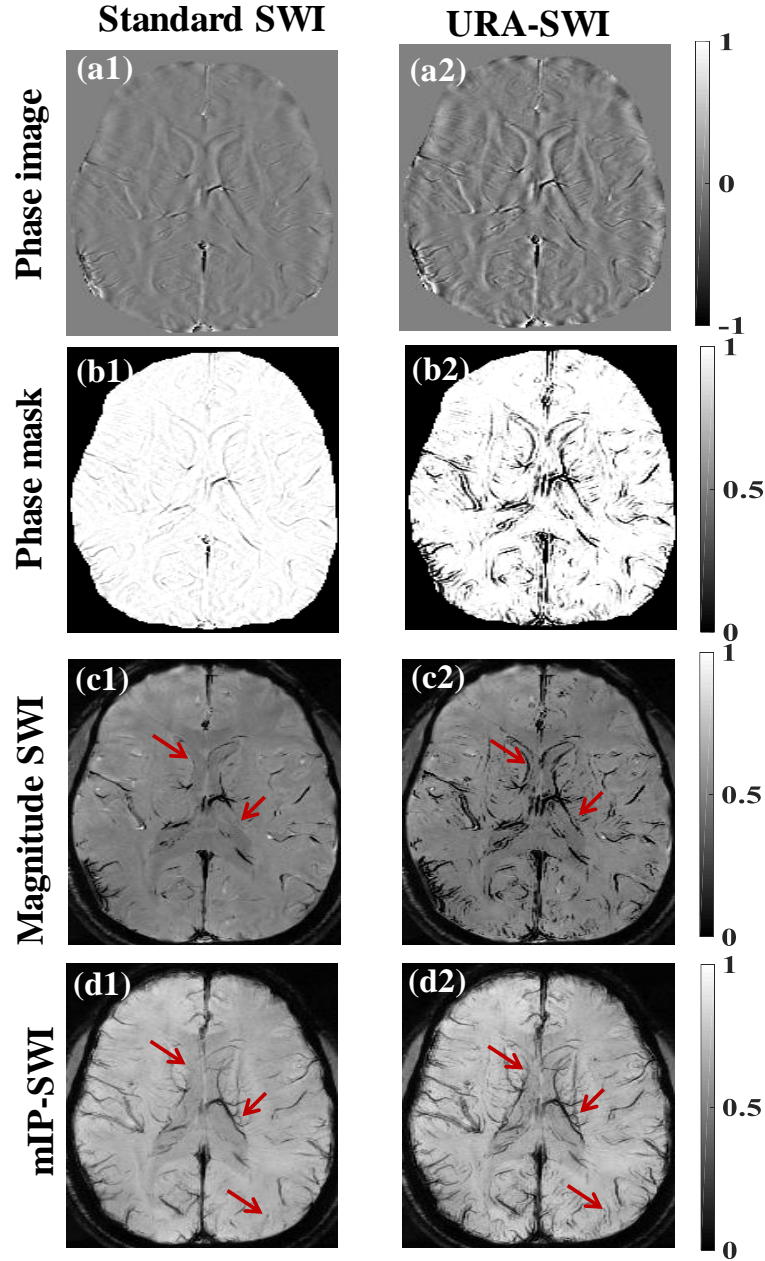

**Figure 6**: Row-wise panels show the combined phase, phase mask, magnitude SWI and mIP images. Red arrows are used to indicate the newly observed venous structures in the reconstructed phase image. These are highlighted only in the corresponding magnitude SWI.

Because of this, the nearby venous structures are obscured in standard SWI. The key message here is that URA processing reduces the remnant effects of non-local field variations, and thereby restores the SuR information. For example, in a brain slice with dural sinus where phase wraps and remnant non-local field inhomogeneities are usually high, phase reconstruction with URA prior enhances the SuR information, and enables restoration of venous structures in the cerebral peduncle and choroid plexus as shown in Fig. 7 (red arrows). Due to the non-local field variations introduced by the interpeduncular cistern, these venous structures are usually not enhanced in standard SWI(Fig.7 (a)). Venous structures between the superior and middle temporal gyrus near to the sinus regions are also seen enhanced in the magnitude SWI generated using the reconstructed phase (red arrow).

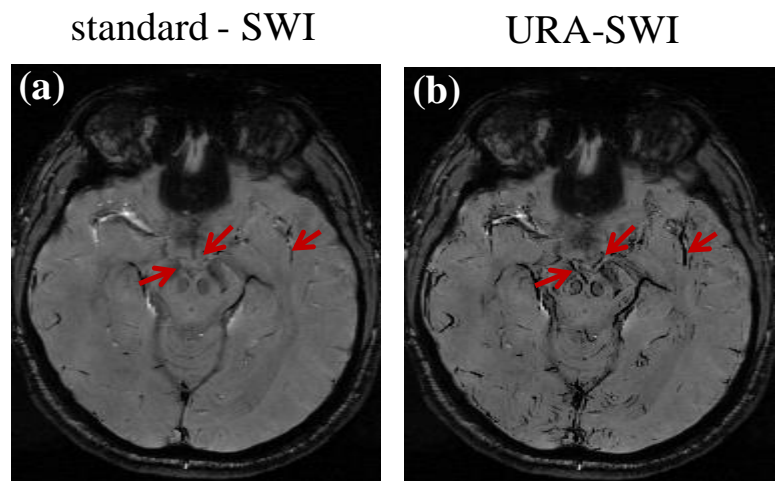

**Figure 7:** Left and right panels show the magnitude SWI generated from the unprocessed phase image and phase image reconstructed with URA prior in a brain slice with dural sinus. Venous structures in the cerebral peduncle and choroid plexus (red arrows) which are obscured in standard SWI are restored in the magnitude SWI generated from the reconstructed phase image.

Contrast enhanced magnitude images are also compared with the susceptibility map weighted imaging (SMWI) and the standard SWI. In SMWI, QSM based weighting provides an alternative SuR contrast to the magnitude images. Left-to-right panels in Fig. 8 show the reconstructed magnitude image, standard magnitude SWI, SMWI and URA-SWI. For contrast enhancement in all three cases, the susceptibility mask is multiplied 4 times with the magnitude image. Among the three methods, standard SWI exhibits comparatively less contrast. Although globuspallidus (GP) appears more hypointense in SMWI, several fine structures are missing in SMWI. Red arrows indicate the subcortical and superficial medullary veins that are missed out in SMWI, and seen with good resolution in the SWI images generated from the reconstructed phase.

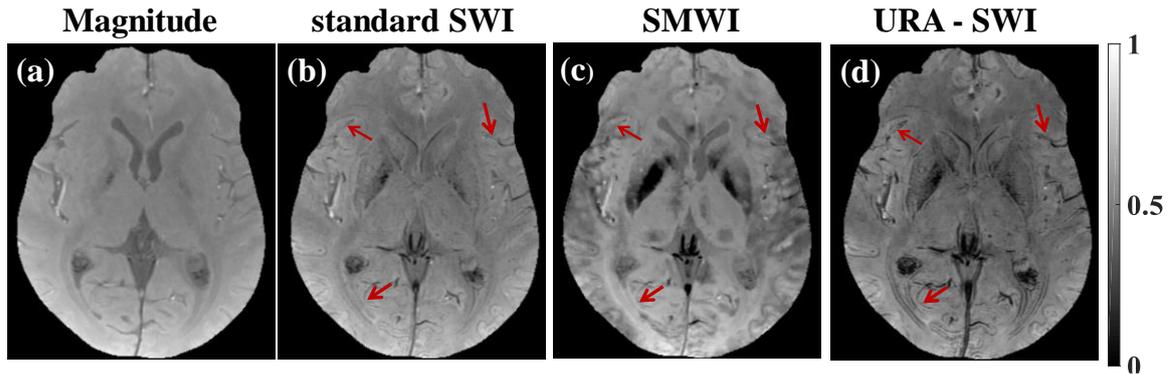

**Figure 8:** Contrast enhanced magnitude SWI generated usingphase image reconstructed with URA prior is compared with standard SWI and SMWI. Left to right panels show the reconstructed magnitude image, standard SWI, SMWI and URA-SWI. Red arrows are used to indicate the sub cortical and superficial medullary veins that are missed out in standard SWI and SMWI.

### 4.2 Phase processing applied to in vivo QSM

Top row in Fig. 9shows the bar plot of average susceptibility in ppm computed from unprocessed phase images, and phase images reconstructed with URA prior. Bar plots show the susceptibility values measured in ROIs that include GP, putamen (PT), caudate nucleus (CN), substantia nigra (SN), red nucleaus (RN),white matter (WM), thalamus (TH) and dentate nucleus (DN). Due to the inherent non-paramagnetic nature, WM and TH regions do not exhibit any significant change in the QSM values computed from the reconstructed phase. This shows that the changes in QSM computed using reconstructed phase are minimal in structures without mineral depositions. It is interesting to note from Fig.9that in iron rich nucleus like GP, PT, CN, SN, RN and DN, QSM computed from the reconstructed phase shows an average enhancement of 40% in the computed susceptibility values. The Dentate nucleus which is a deep cerebellar nucleus, shows the highest susceptibility variations from 0.0442±0.0201 to 0.1071±0.0406. Furthermore, QSM values of SN and RN nuclei, located in regions with high phase wraps and background field, are enhanced by 22.6% and 35.2% with respect to the standard QSM. Bottom three rows show the susceptibility maps computed from the unprocessed phase (left) and reconstructed phase (right). Red arrows indicate the newly observed SuR features due to iron depositions in insular, motor cortex and superior frontal gyrus.

## 5. DISCUSSION

In this study, we use a regularized phase reconstruction scheme with URA prior for restoration of venous structures, especially in regions with high phase wraps and macroscopic field variations. Although several algorithms have been developed to unwrap the phase by enforcing spatial smoothness, the imperfections in phase unwrapping will introduce errors leading to loss of SuR information, and hence causing underestimation of susceptibility values in regions with high phase wraps. The same applies to the algorithms used for background suppression also. In particular, regions around focal susceptibility sources such as highly paramagnetic cerebral microbleeds are most vulnerable to the unwrapping induced errors(22). This is because the dipolar field perturbations due to susceptibility variations will result in proportionate amounts of phase accumulation, thereby introducing phase wraps that exist below the resolution of the image, and hence unresolved in the unwrapped phase. Further, the incoherence in gradient refocusing due to presence of macroscopic field variations will lead to phase dispersion across voxels, resulting in magnitude intensity dropout. With reference to the signal model in the temporal dimension, the associated sinc function modulation makes the magnitude component unpredictable. Furthermore, the phase SNR being directly related to the magnitude signal intensity, the phase images will have low SNR at locations with signal dropout. This effect will be more severe in the regions influenced by the macroscopic field. Due to the imperfections in background suppression, phase dispersion is not completely eliminated from the phase image. The combined effect of reduced SNR and phase dispersion causes the voxel-wise signal model to deviate from the one consisting of a single complex exponential corresponding to the frequency shift introduced by the susceptibility induced field. Thus by model fitting the unit magnitude complex exponential in the background suppressed phase image, it is possible to apply a unity rank approximation of the structured Hankel matrix along the temporal dimension to reduce the influence of phase dispersion.

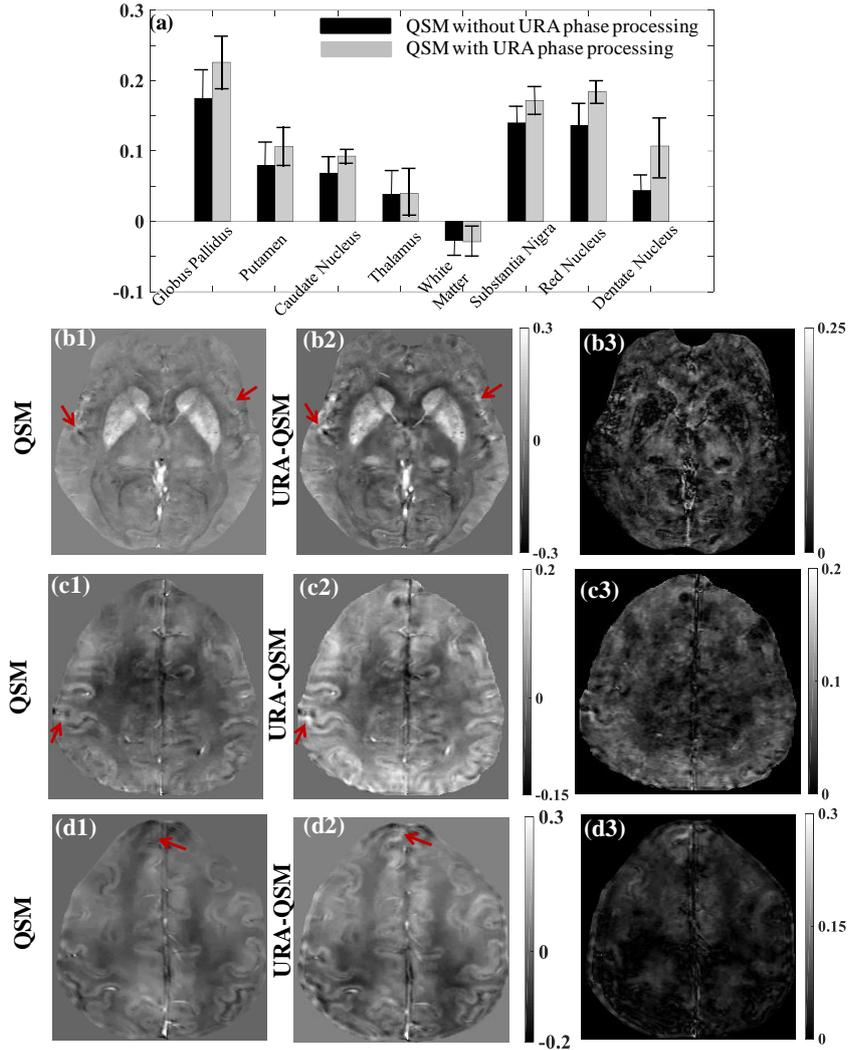

**Figure 9:** Top row shows the bar plot of average susceptibility in ppm computed from the unprocessed phase image and phase image reconstructed with URA prior. Bar plots show the susceptibility values measured in ROIs that include globus-pallidus (GP), putamen (PT), caudate nucleus (CN), substantia nigra (SN), red nucleus (RN), white matter (WM), thalamus (TH) and dentate nucleus (DN). Bottom three rows show the susceptibility maps computed from the unprocessed phase (left) and reconstructed phase (right). Red arrows in panel (b)-(d) indicate the newly identified SuR information in URA-QSM signifying iron depositions in the insular, motor cortex and superior frontal gyrus. The last column shows the difference images computed between the QSM maps generated without and with URA priors.

Non-linear evolution of phase was earlier reported byWharton and Bowtell (56) in healthy brain regions where the local signal phase can be correctly recovered via unwrapping and background field removal. In addition, several research groups have attributed other reasons that lead to non-linear phase evolution in GRE. This include exchange processes(57), NMR invisible microstructures(58) and anisotrpic magnetic susceptibility effects(59). It should be remarked that application of regularized phase reconstruction with URA prior, fails to fit the extra frequency components that model these non-linear effects. The MSE measures indicated by our simulation studies model only the linear phase evolution using a phantom with isotropic susceptibility distribution. The MSE computations reflect only the direct Fourier relationship between the isotropic susceptibility distribution and the dipolar field(60).

The primary to the secondary singular value magnitude ratio of the unprocessed Hankel matrix at each voxel act as an indicator of the effect of phase dispersion. The effect of phase dispersion is higher at regions where this ratio becomes closer to unity. It is observed that with reconstruction using URA prior, this ratio reduces, that implicitly shows the diminishing effect of phase dispersion. Also, closeness of the non-principal singular values to zero is indicative of the effectiveness of background suppression. Upon reconstruction with URA prior, the non-principal singular values are effectively forced to zero, indicative of the fact that the reconstruction procedure effectively reduces the sensitivity of further processing steps such as the susceptibility estimation, on the method employed for background suppression. Further numerical evidence for the aforementioned facts are provided in the supplementary material annexed to this article.

**CONCLUSIONS**

This work proposes incorporation of URA prior with data fidelity into the SWI processing pipeline to enable restoration of structures that act as potential susceptibility sources, and the QSM pipeline for more robust estimation of the underlying susceptibility distribution. Using in vivo data, it is demonstrated that our phase reconstruction method enables restoration of venous structures in cerebral peduncle and temporal gyrus that are obscured in magnitude SWI generated with the state-of-the-art methods. It is also shown that QSM computed from the reconstructed phase image exhibit new SuR features and more accurate susceptibility values in the iron rich nuclei.

10. Deistung A, Schweser F, Wiestler B, Abello M, Roethke M, Sahm F, et al. Quantitative susceptibility mapping differentiates between blood depositions and calcifications in patients with glioblastoma. PloS one. 2013;8(3):e57924.

11. Rauscher A, Sedlacik J, Barth M, Mentzel H-J, Reichenbach JR. Magnetic susceptibility-weighted MR phase imaging of the human brain. American Journal of Neuroradiology. 2005;26(4):736-42.

12. Lee J, Hirano Y, Fukunaga M, Silva AC, Duyn JH. On the contribution of deoxy-hemoglobin to MRI gray–white matter phase contrast at high field. Neuroimage. 2010;49(1):193-8.

13. Madhusoodhanan S, Kesavadas C, Paul JS. SWI processing using a local phase difference modulated venous enhancement filter with noise compensation. Magnetic resonance imaging. 2019;59:17-30.

14. Gho SM, Liu C, Li W, Jang U, Kim EY, Hwang D, et al. Susceptibility map-weighted imaging (SMWI) for neuroimaging. Magnetic resonance in medicine. 2014;72(2):337-46.

15. Helms G, Dechent P. Increased SNR and reduced distortions by averaging multiple gradient echo signals in 3D FLASH imaging of the human brain at 3T. Journal of Magnetic Resonance Imaging: An Official Journal of the International Society for Magnetic Resonance in Medicine. 2009;29(1):198-204.

16. Denk C, Rauscher A. Susceptibility weighted imaging with multiple echoes. Journal of Magnetic Resonance Imaging. 2010;31(1):185-91.

17. Zhou K, Zaitsev M, Bao S. Reliable two-dimensional phase unwrapping method using region growing and local linear estimation. Magnetic Resonance in Medicine: An Official Journal of the International Society for Magnetic Resonance in Medicine. 2009;62(4):1085-90.

18. Reichenbach J, Schweser F, Serres B, Deistung A. Quantitative susceptibility mapping: concepts and applications. Clinical neuroradiology. 2015;25(2):225-30.

19. Lee J, Nam Y, Choi JY, Kim EY, Oh SH, Kim DH. Mechanisms of T2* anisotropy and gradient echo myelin water imaging. NMR in Biomedicine. 2017;30(4):e3513.

20. Funai AK, Fessler JA, Yeo DT, Olafsson VT, Noll DC. Regularized field map estimation in MRI. IEEE transactions on medical imaging. 2008;27(10):1484-94.

21. Dagher J, Reese T, Bilgin A. High-resolution, large dynamic range field map estimation. Magnetic resonance in medicine. 2014;71(1):105-17.
29